\documentclass{article}

\usepackage[hmarginratio=1:1,top=28mm,left=20mm,right=20mm,bottom=28mm,columnsep=9mm]{geometry} 
\usepackage{multicol} 

\usepackage{booktabs} 

\usepackage{lettrine} 
\usepackage{paralist} 

\usepackage{abstract} 

\usepackage{titlesec} 

\usepackage{amsmath}
\usepackage{amsfonts}
\usepackage{amssymb}

\usepackage{upgreek}
\usepackage{graphicx}
\usepackage{natbib}
\usepackage[normal,small,bf,figurename=Fig.]{caption}

\usepackage{tikz}

\newcommand{\figurewidthsingle}{74mm}
\newcommand{\figurewidthdouble}{175mm}

\newenvironment{Figure}
  {\par\medskip\noindent\minipage{\linewidth}}
  {\endminipage\par\medskip}
	
\newenvironment{Table}
  {\par\medskip\noindent\minipage{\linewidth}}
  {\endminipage\par\medskip}

\title{\vspace{-15mm}{Hand-guided 3D surface acquisition by combining simple light sectioning with real-time algorithms}}
\author{Oliver Arold $\cdot$ Svenja Ettl $\cdot$ Florian Willomitzer $\cdot$ Gerd H\"{a}usler \smallskip
\\ \small University Erlangen-Nuremberg \\ \small Institute of Optics, Information and Photonics \\ \small 91058 Erlangen, Germany}
\date{}

\begin{document}

\maketitle 



\begin{abstract}

\noindent 
Precise 3D measurements of rigid surfaces are desired in many fields of application like quality control or surgery. Often, views from all around the object have to be acquired for a full 3D description of the object surface.
We present a sensor principle called ``Flying Triangulation" which avoids an elaborate ``stop-and-go" procedure. It combines a low-cost classical light-section sensor with an algorithmic pipeline. A hand-guided sensor captures a continuous movie of 3D views while being moved around the object. The views are  automatically aligned and the acquired 3D model is displayed in real time.
In contrast to most existing sensors no bandwidth is wasted for spatial or temporal encoding of the projected lines. Nor is an expensive color camera necessary for 3D acquisition.
The achievable measurement uncertainty and lateral resolution of the generated 3D data is merely limited by physics.
An alternating projection of vertical and horizontal lines guarantees the existence of corresponding points in successive 3D views. This enables a precise registration without surface interpolation. 
For 
registration, a variant of the iterative closest point algorithm -- adapted to the specific nature of our 3D views -- is introduced. 
Furthermore, data reduction and smoothing without losing lateral resolution as well as the acquisition and mapping of a color texture is presented. 
The precision and applicability of the sensor is demonstrated by simulation and measurement results.

\end{abstract}


\begin{multicols}{2} 

\section{Introduction}
\label{sec_introduction}

Optical sensors allow a fast, contact-free 3D acquisition of surfaces and become more and more the number-one choice for 3D metrology: Industry applies optical sensors for rapid prototyping and quality control. In the medical field 3D measurements provide crucial pre-, intra-, and post-operative information for the surgeon. Intra-oral sensors enable a comfortable way of  manufacturing tooth crowns. In the field of cultural heritage optical 3D sensors are a helpful tool for restoration, documentation, and duplication of sculptures.  

Most of these applications require 3D views from many different directions to cover the relevant surface of the rigid object under test. This leads to an elaborate and time consuming repositioning of the sensor or the object (see Figure~\ref{fig1}). Furthermore, the acquired 3D views have to be aligned to each other to yield a complete 3D model. This alingment can be done by external tracking systems or by registration algorithms. In the first case, the movement of the sensor is commonly restricted by the tracking system. In the second case, enough overlap and common information between the separate 3D views must be guaranteed.


We present a sensor principle which avoids a ``stop-and-go" measurement procedure. Instead, a simple multi-line light-section sensor acquires a continuous series of 3D views while being hand guided around the object. Overlap between consecutive views is guaranteed. No tracking is necessary for an automatic alignment of the views. 
A key characteristic of the proposed sensor compared to other hand-guided systems is that no bandwidth is wasted for spatial or temporal encoding methods. 
The width of each projected line is as narrow as possible and as wide as necessary to allow a precise subpixel localization. 
This leads to both high depth and high lateral resolution of each calculated 3D point. But in return 3D points are only acquired along few separate lines.
To our knowledge, most other hand-guided sensor systems abandon measurement accuracy in order to obtain more dense 3D views and hereby allow the application of well-established surface or dense-point-cloud registration methods for the alignment of the views. 

\begin{Figure}
 \centering
 \includegraphics[width=\figurewidthsingle]{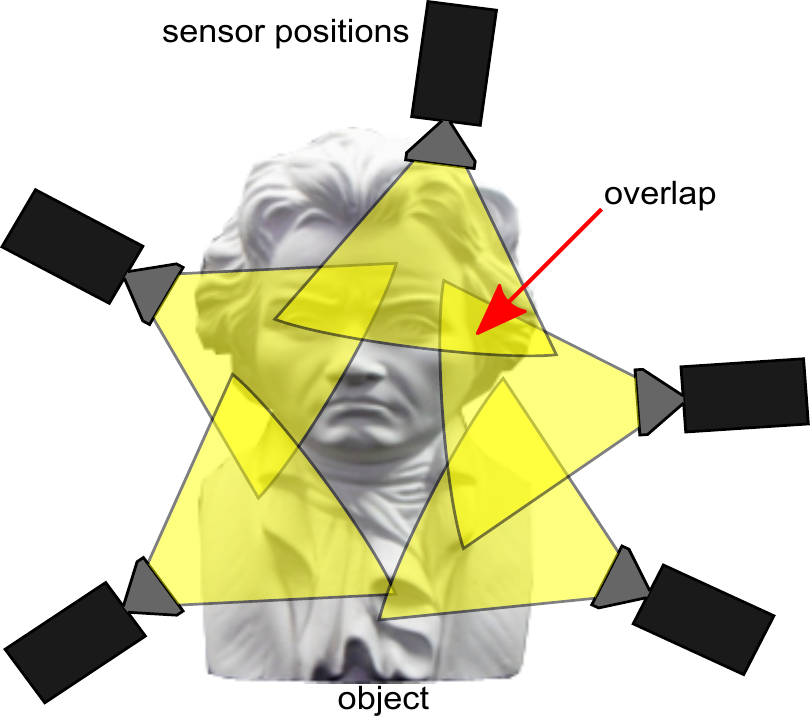}
 \captionof{figure}{An all-around measurement requires an acquisition of views from many different directions}
 \label{fig1} 
\end{Figure}

In our case, a robust and precise real-time registration of the acquired series of sparse 3D views is enabled by an alternating projection of orthogonal line patterns and by applying a specifically developed algorithmic pipe\-line.
There is no need for surface interpolation across the acquired line profiles. Furthermore, an on-time visualization of the current measurement result provides the operator with information about not yet acquired object areas.

The physical setup and optimization of the underlying light-section sensor and the basic measurement principle are described in \citet{Ettl12} and summarized in Section~\ref{measurement_principle}.
The main focus of the present paper is a description of the algorithmic architecture, which is the most crucial part of the suggested sensor principle. This comprises 3D data generation, line indexing, registration, visualization, data reduction, and texture mapping. 
Eventually, the sensor performance is evaluated by simulations and measurements.

\section{Related Work}
\label{sec_related_work}

This section gives an overview of other hand-guided sensor systems and registration algorithms. 
The key characteristics are described and compared. 
Furthermore, the weaknesses that are overcome by the approaches in this paper are identified.

\subsection{Hand-guided Sensors}
\label{sec_other_sensors}

The focus in this section lies on sensors and techniques that enable all-around measurements of static objects. Methods for the acquisiton or reconstruction of dynamic scenes aim for different kinds of applications and thus are not discussed~\citep{Koninckx03,Kawasaki08}.\\

The \textit{``Artec 3D Scanner"} is a hand-guided sensor without need for external tracking~\citep{Suhovey10}. It is based on light sectioning and uses a spatially encoded line pattern to increase the number of distinguishable lines. A continuous stream of 15 views per second is acquired. Due to the relatively high data density standard full-field registration methods can be applied to automatically align the captured views in real time. Colored texture can be captured and mapped onto the data. Since the lines are spatially encoded both the lateral and longitudinal resolution are reduced and small details are not resolved.

Rusinkiewicz also describes a hand-guided sensor system based on light section\-ing with an encoded line pattern~\citeyearpar{Rusinkiewicz02}. In this case the pattern is temporally encoded and projected onto the object by DLP technology. From each single camera image 3D data is calculated, but four consecutive images are necessary in order to find the correct line indices. This leads to a restriction of both sensor movement and object topography: The sensor has to be moved slowly and the object must not comprise high steps. A variant of the iterative closest point algorithm is applied to align the views in real time.

A further hand-guided sensor based on light section\-ing is described by Matabosch et al~\citeyearpar{Gerones07,Matabosch08}. A continuous stream of 15 views per second is acquired and registered in real time. Each view contains 3D data along 19 separate lines. The correct indexing of the uncoded lines requires the visibility of all lines in at least one row of the camera image. This considerably restricts both the object surface and the movement of the sensor. 
Furthermore, the accuracy of the applied registration is limited due to the need for surface interpolation between the acquired lines.

The \textit{``kolibri CORDLESS"} is a hand-held sensor system based on fringe projection \citep{Kuemstedt07}. For a more robust and  error-tolerant phase evaluation a stereo approach is applied. Since a series of camera images is necessary to calculate 3D data both sensor and object are not allowed to move during the acquisition of one 3D view. This makes the system motion sensitive, but in return each view provides full-field 3D data. Instead of a continuous stream, views from distinct positions and directions are acquired. Due to the employment of a fast projection unit with at least 60 fps, the sensor can be hand held while acquiring one single view. To enable an automatic registration the operator has to ensure sufficient overlap between the acquired views.

As described above, we aim for a sensor that renders none of these drawbacks: no waste of bandwidth by line encoding -- and thus no sacrifice of lateral resolution, no restriction of sensor movement for correct line indexing, no ``stop-and-go" procedure, and no surface interpolation to enable registration.

\subsection{Registration}
\label{sec_other_algorithms}

The alignment of two 3D views is commonly split into two steps. First, a coarse registration finds an initial transformation between the views. Then, a fine registration improves the result by an iterative procedure. A general overview over common registration techniques is given by \citet{Seeger00} and \citet{Salvi07}.

Most coarse registration techniques are based on the detection of common features in the overlapping region of the views. Therefore, 3D neighborhood information is necessary. \citet{Schon06} search for so-called salient points in both views and map them onto each other. A method especially developed for technical surfaces with planar parts is presented by \citet{Maier06}. \citet{Johnson97} and \citet{Kaminski07} present approaches suitable for free-form surfaces. For \citet{Faugeras86} and \citet{Stein92} lines, surface parts or objects serve as features and are used for an initial mapping. 
Gelfand calculates descriptor values based on local geometry to identify common features in partially overlapping views \citeyearpar{Gelfand05}.
   
For the fine registration of two point clouds most often methods based on the iterative closest point algorithm (ICP) are used. The ICP was first introduced by Besl and Chen \citeyearpar{Besl92,Chen92}. Further improvements were given by Zhang \citeyearpar{Zhang94}. A comprehensive overview is given by \citet{Rusinkiewicz01}. The key idea of the ICP algorithm is to find corresponding points in two overlapping views. In each iteration the points closest to each other are assumed to correspond and the best mapping transformation in terms of least squares is calculated. The accuracy of the registration result mainly depends on the precision and local curvature of the acquired 3D data \citep{Laboureux01}.

In our case, none of these methods can be applied directly, since we only acquire 3D information along few separated line profiles. For this reason a variant of the ICP specifically adapted to the nature of our sparse 3D views is presented in Section~\ref{sec_registration}.


\section{Flying Triangulation}
\label{sec_sensor}

In this section, the measurement principle, the sensor setup, and the physical optimization of Flying Triangulation is described. 

\subsection{Measurement Principle}
\label{measurement_principle}

\begin{figure*}[!t]
	\centering
	\includegraphics[width=\figurewidthdouble]{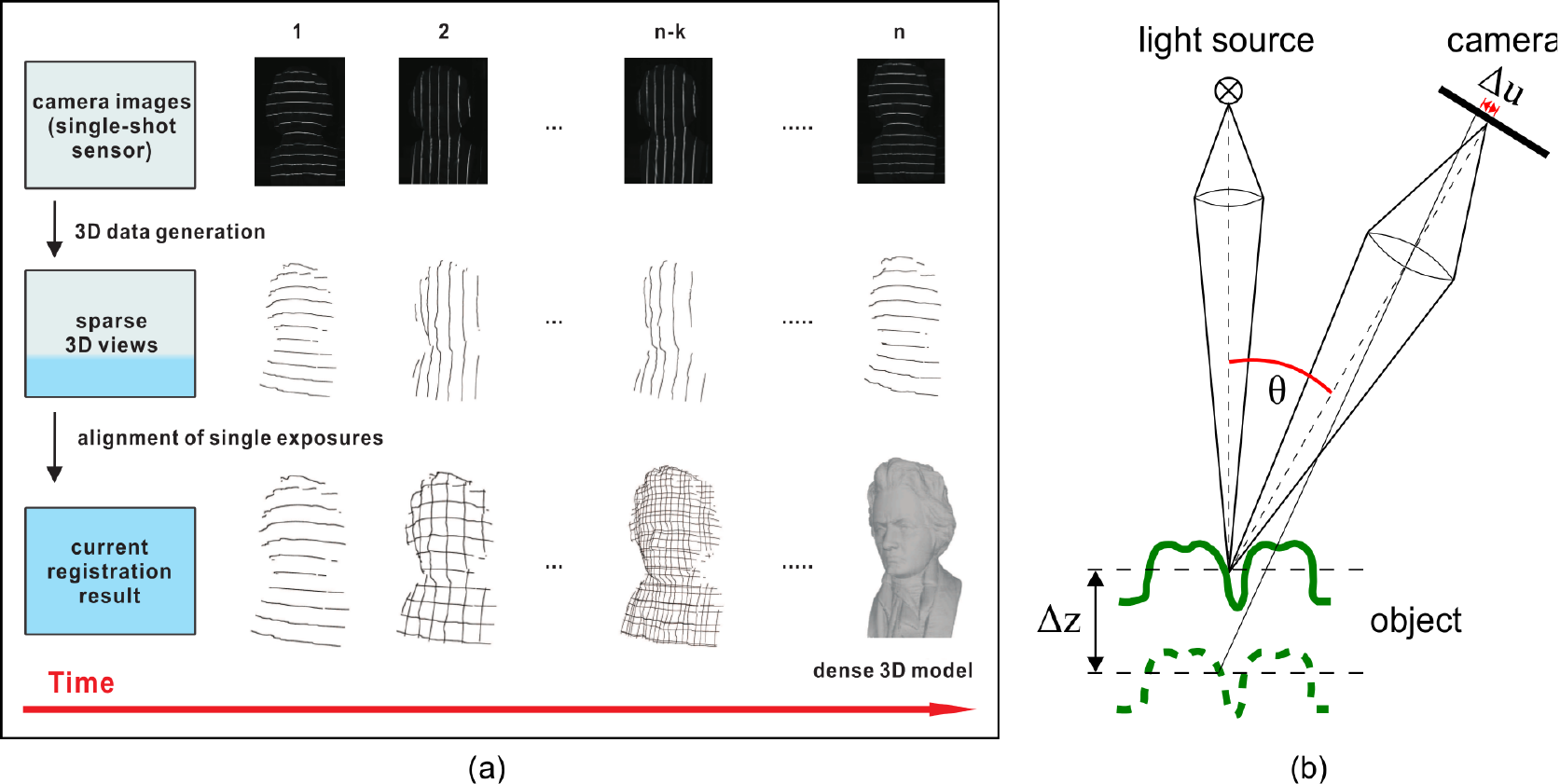}
	\caption{(a) Flying Triangulation principle. A multi-line single-shot sensor generates a series of sparse 3D views. The views are aligned to each other in real time  to yield a complete and dense 3D surface model with high lateral resolution. (b) The single-shot sensor is based on triangulation. A line pattern is projected onto the object and observed by a camera under a triangulation angle $\theta$. From the displacement $\Delta u$ of the line in the camera image the depth $\Delta z$ can be calculated}
\label{fig2}      
\end{figure*}

The basic principle is depicted in Figure~\ref{fig2} (a). 
Two line patterns, which are oriented perpendicular to each other, are alternately projected onto the surface and observed by a camera with at least 30 frames per second. 
From each camera image a sparse 3D view consisting of line profiles is generated. The views are aligned in real time and the visualization of the current result serves as feedback to the operator.

\subsection{Sensor Setup}
\label{sec_setup}

The single-shot sensor is based on the well-known light sectioning illustrated in Figure~\ref{fig2} (b): Lines are projected onto the surface of the object and observed by a camera under a triangulation angle $\theta$. From the displacement $\Delta u$ of the lines in the camera image the depth $\Delta z$ is calculated. This calculation is done by using the calibration technique described in Section~\ref{sec_calibration}. No spatially or temporally encoded pattern is employed. As a result, each acquired 3D point provides high precision and high lateral resolution, as shown below. 

\begin{figure*}[!t]
	\centering
	\includegraphics[width=\figurewidthdouble]{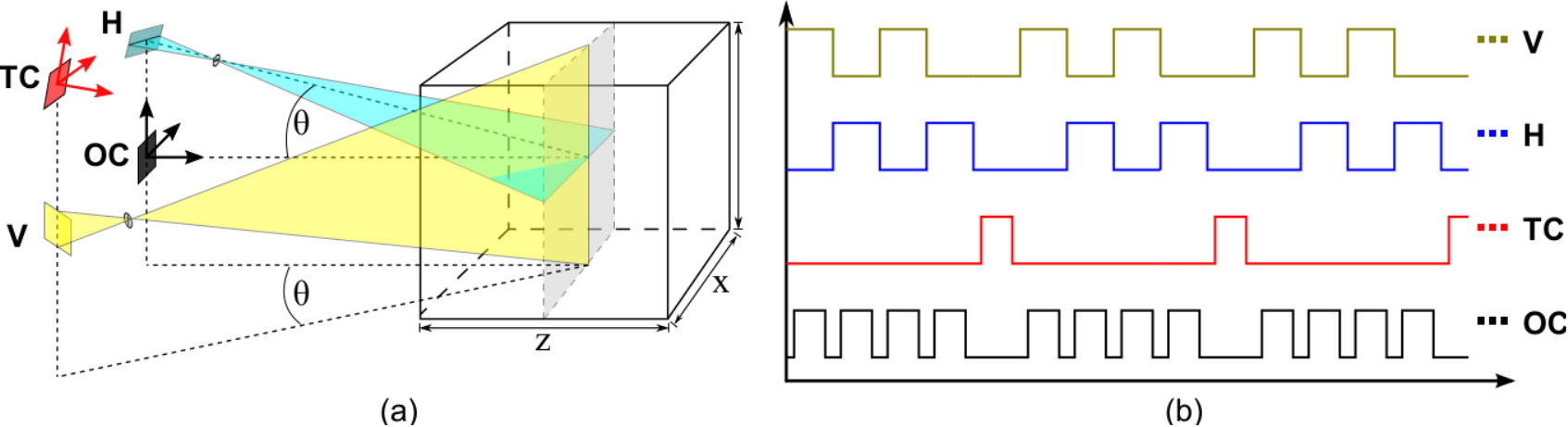}
	\caption{(a) Basic hardware setup. Two illumination units (V and H) alternately project vertical and horizontal lines into the measurement volume. One observation camera (OC) acquires a continuous series of images for the 3D data generation. A texture camera (TC) captures colored texture images. (b) Sketch of the triggering signals for the illumination units and cameras}
\label{fig3}      
\end{figure*}

Since there is a considerable spacing between the line profiles, no dense surface information is obtained in one single 3D view. Therefore, a trick is applied, which makes an accurate registration of subsequent sparse views possible: 
Instead of one single projection unit, two 
projection units are integrated in the setup (see Figure~\ref{fig3}). One projects a vertical and the other a horizontal line pattern. A detailed description of the registration algorithm, specifically developed for this purpose, is discussed in Section~\ref{sec_registration}.

\subsection{Physical Optimization}
\label{sec_optimization}

In order to achieve the highest possible precision the setup of the sensor is physically optimized. The key aspects and parameters that have to be taken into account are summarized. A more detailed discussion is given in \citep{Ettl12}. \\

According to \citet{Dorsch94} the fundamental cause of measurement uncertainty for a light-section sensor is speckle noise:

\begin{equation}
\label{speckle}
	\delta z = \frac{C}{2\pi} \, \frac{\lambda}{\sin u_{\rm obs} \, \sin \theta},
\end{equation} 
where $C$ is the speckle contrast, $\lambda$ the wave length, $\sin u_{\rm obs}$ the observation aperture, and $\theta$ the triangulation angle. Unfortunately, most of the parameters cannot be chosen independently. Thus, trade-offs have to be made to minimize the measurement uncertainty within the entire desired measurement volume.\\
A high observation aperture $\sin u_{\rm obs}$ reduces the speckle noise, but also decreases the depth of focus and thereby leads to a higher out of focus measurement uncertainty.\\
Since the sensor should be able to measure deep gaps without being restricted by shadowing effects, the triangulation angle is limited. Experiments have shown that for most applications a triangulation angle of about $7^{\circ}$ is a good compromise.

For a motion-robust acquisition the exposure time $t_{\mbox{\small exp}}$ for a single 3D view should be below 30ms. Thus, light source and apertures have to be chosen to provide sufficient light. Again, one has to bear in mind that larger apertures reduce the depth of focus.

The remaining option to minimize speckle noise is the reduction of the speckle contrast $C$. For this purpose, the following strategies are applied. 
First, no lasers are employed. Instead, white-light LEDs and lithographic patterns are used for the projection of the lines. The low temporal coherence of the white light helps to reduce the temporal speckle contrast in the presence of volume scattering. This way, the signal-to-noise ratio can be improved by almost a factor of 10 compared to laser illumination. If necessary, systematic height errors due to strong volume scattering can be reduced by spraying the surface with titanium dioxide. The thickness of the sprayed layer is about 20$\upmu$m. Since its surface roughness is still larger than the coherence length of the white-light source, the speckle contrast is still low.
Furthermore, for a reduction of the spatial coherence the setup comprises an illumination aperture larger than the observation aperture \citep{Haeusler03}. Again, the choice is limited by the desired depth of focus.

\begin{figure*}
	\centering
	\includegraphics[width=\figurewidthdouble]{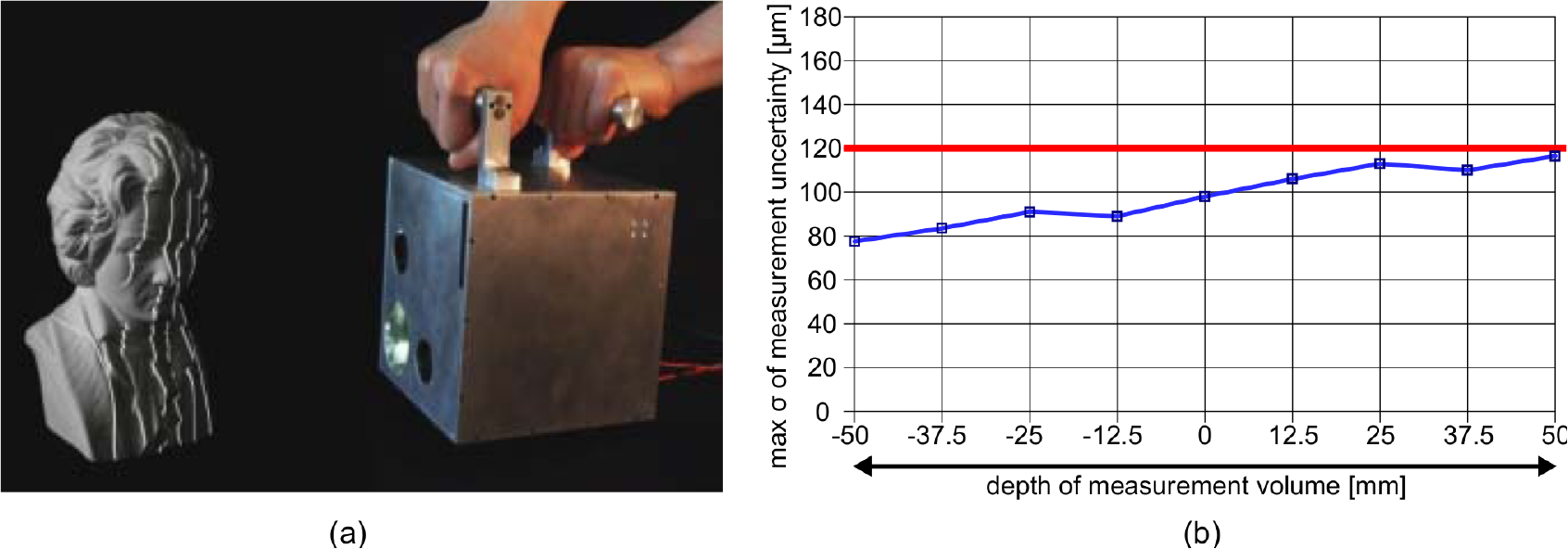}
	\caption{(a) Picture of the sensor during a measurement. (b) Measurement uncertainty in $\upmu$m depending on depth in mm within the measurement volume}
\label{fig4}      
\end{figure*}

Considering all these aspects, we implemented a sensor for objects like faces or sculptures with a measurement uncertainty of less than 120$\upmu$m inside a measurement volume of about 150mm $\times$ 200mm $\times$ 100mm. Figure~\ref{fig4} displays the sensor and the achieved measurement uncertainty along the depth of the measurement volume. \\
For the measurement of (sprayed) teeth we implemented an intra-oral sensor with a measurement uncertainty of less than 30$\upmu$m within a measurement volume of 20mm $\times$ 15mm $\times$ 15mm. \\

It should be noted that besides a small measurement uncertainty in the $z$-direction, the ability to laterally resolve small shape details in 3D space is an important feature of 3D sensors. This will briefly be discussed:

\textit{Perpendicular to the lines}, only lateral details bigger than the width $w_0$ of the projected lines on the object surface can be reliably measured and resolved. 
For example: to measure two grooves oriented in the line direction, these grooves must have a distance (pitch) not smaller than $2w_0$ (we assume properly band limited grooves).
So, for the lateral resolution in a direction perpendicular to the lines, the width $w_0$ should be as small as possible.

However, the pixelated camera image of the projected line, with a width $w'_0$,  must allow a precise subpixel localization of the line signal. This is guaranteed if the camera imaging satisfies the sampling theorem. A proven and tested approach is that the full-width-half-maximum dimension of the line signal at the camera chip covers at least 3 pixels. This corresponds to a minimum total line width  $w'_0 \approx 4p$, where $p$ is the pixel pitch. This optimal line width will allow for the best possible resolution.

In practice, due to the limited depth of focus, the line width at the object varies along the $z$-axis. For the face sensor, the line width ranges from about 800$\upmu$m to 1200$\upmu$m and, for the intra-oral sensor, it varies from about 80$\upmu$m to 180$\upmu$m. This corresponds to a line width $w'_0$ at the camera chip of 4$p$-6$p$, respectively 4$p$-9$p$.

These considerations are valid both for the horizontal and vertical lines. 

We still have to estimate the lateral resolution \textit{along the line direction}. Principally, this is not much different. A properly designed camera should satisfy the sampling theorem in any direction. So the camera image of the projected lines (and, hence, the details of the 3D object structure) is smoothed along the line direction in the same way as across the line direction. 

It remains to be mentioned: If the line width at the camera chip is bigger than the minimal allowed line width 4$p$, the lateral resolution along the line direction is different from the resolution across the line direction: The latter is given by the line width itself (as explained above), while the resolution along the line direction is given by the point spread function of the camera lens. So it may happen that the resolution for grooves perpendicular to the line direction is better than the resolution across the line direction. 

To conclude: For the best possible lateral resolution, the width of each projected line should be as small as possible. At the same time, the width of the line signals at the camera chip must not be smaller than allowed by the sampling theorem, in order to achieve the minimal measurement uncertainty in the $z$-direction. 

Table 1 summarizes the most important specifications of the face sensor stemming from physical optimization. More details about the intra-oral sensor can be found in~\citet{Ettl12}.

%
				%
			%

\begin{Table}	    
	\centering
	\captionof{table}{Physical specifications of the face sensor.}
	\label{tab_sensor_face}	

	\begin{tabular}{|p{4.5cm}|p{3.0cm}|} 
		\hline\noalign{\smallskip}
				
		Light source 												&	White-light LED									\\
		Measurement volume 									& 150$\times$200$\times$100mm$^3$	\\	
		Triangulation angle 								& 7$^{\circ} $										\\
		Vertical Lines 											& 12  														\\
		Horizontal Lines 										& 9  															\\
		Measurement uncertainty							& $<$ 120$\upmu$m  								\\
		Pixel resolution										& $\approx$ 200$\upmu$m  					\\
		Line width 						              & 800$\upmu$m-1200$\upmu$m 				\\	
			
		\noalign{\smallskip}\hline	
	\end{tabular}	
\end{Table}

\section{3D Data Generation}
\label{3d_data_generation}

In this section the calibration and the challenges of calculating correct and precise 3D data from each individual camera image are described. 

\subsection{Line Localization}
\label{sec_localization}

As mentioned in Section~\ref{sec_setup}, for light sectioning 3D data is calculated from the positions of the lines in the camera images. Thus, the measurement uncertainty mainly depends on the localization of these lines. 
Since the observed images are pixelated, an interpolation has to be used to calculate subpixel-precise localizations. We apply the well-known three point Gaussian interpolation. This is a reasonable choice given the intensity profile of a line being similar to a Gaussian curve. 
Figure~\ref{fig5} illustrates the interpolation and localization process.
Additionally, in order to reduce the influence of noise, the intensity profiles of the lines are smoothed by a narrow Gaussian filter before the calculation.
The localization is performed for each row (in case of vertical lines) or each column (in case of horizontal lines) of the camera image. With the proposed method and real noise a statistical localization uncertainty of about one seventh of a pixel has been achieved which corresponds to the achieved measurement uncertainty of $\delta z \approx 120\upmu$m (face sensor) and $\delta z \approx 30\upmu$m (intra-oral sensor).


\begin{Figure}
 \centering
 \includegraphics[width=\figurewidthsingle]{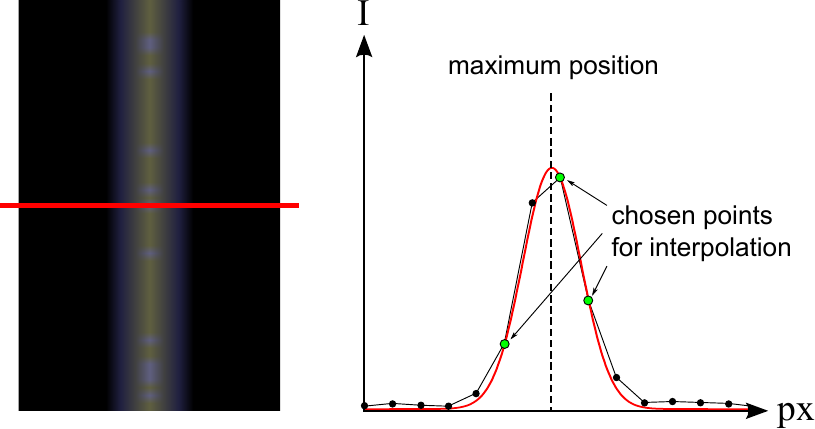}
 \captionof{figure}{A Gaussian bell curve is fitted to the intensity profile in order to localize the line with subpixel precision}
 \label{fig5} 
\end{Figure}

\subsection{Calibration} 
\label{sec_calibration}

The calibration of the sensor system enables the calculation of 3D data from the line position obtained by the localization process described above. Most established calibration methods for camera-projector systems perform a model-based intrinsic calibration of the camera but lack a corresponding intrinsic calibration of the projector. The distortion of the projected pattern caused by the imaging system of the projector is neglected. 
For the light-sectioning sensor proposed in this paper a model-free calibration is applied which incorporates the intrinsics of the imaging systems of camera and projector. 
The methhod can be split into two parts: a $z$-calibration for the calculation of depth values and an $xy$-calibration to obtain the right metric along the $x$- and $y$-axis. The $z$-calibration has to be calculated for both the vertical and horizontal line projection.

\subsection{$z$-Calibration} 
\label{sec_calibration_z}

The $z$-calibration procedure is explained for the vertical line pattern only, since the horizontal calculation is basically the same. 

\begin{figure*}
	\centering
	\includegraphics[width=\figurewidthdouble]{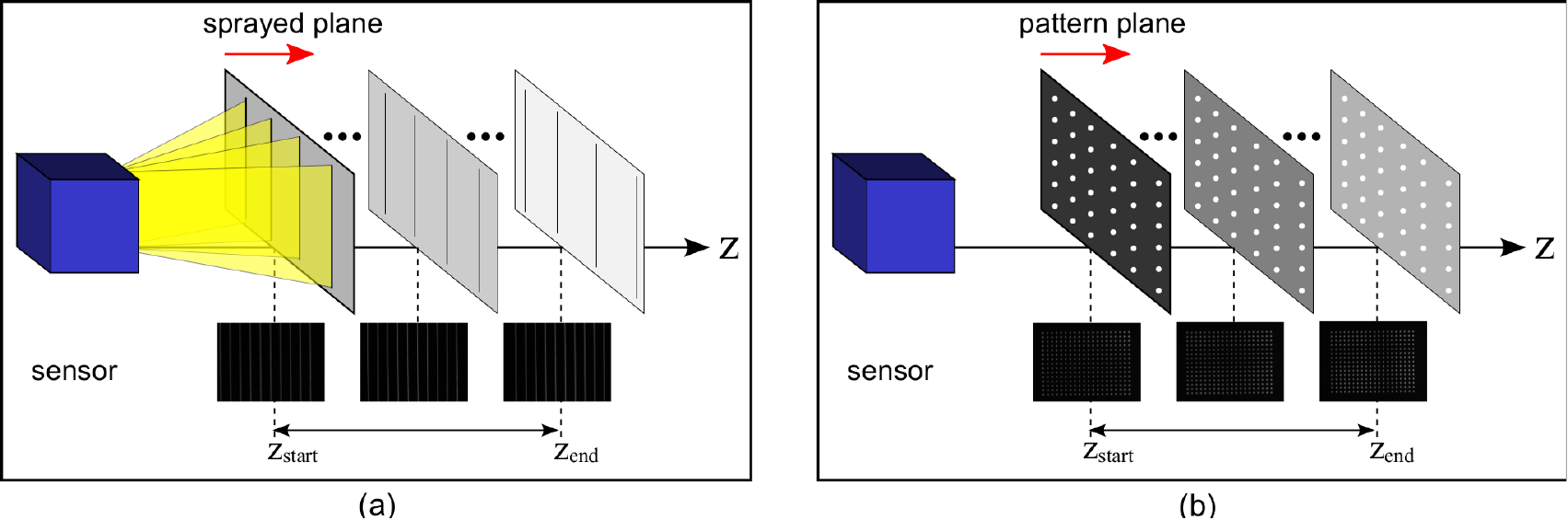}
	\caption{(a) $z$-Calibration: A sprayed planar mirror is moved along the entire measurement depth. At known $z$-positions between $z_{\mbox{\small start}}$ and $z_{\mbox{\small end}}$ camera images with the projected line patterns are acquired.  (b) $xy$-Calibration: A plane with a spot pattern is moved along the entire measurement depth. At known $z$-position between $z_{\mbox{\small start}}$ and $z_{\mbox{\small end}}$ camera images of the spot pattern are acquired}
\label{fig6}      
\end{figure*}

Goal of the proposed $z$-calibration is the determination of polynomial transformations $Z_{v}^{n}$ with

\begin{equation} 
	z = Z_{v}^{n}(u) = a_{v}^{n} + b_{v}^{n} u + c_{v}^{n} u^2 + d_{v}^{n} u^3,
	\label{e_calib_z}
\end{equation}
where $z$ is the desired depth value, $n$ is the line number, $v$ the row of the camera image and $u$ is the subpixel-precise position of the line signal in the camera image (see Section~\ref{sec_localization}). 
For every line $n$ and every row $v$ the set of polynomial coefficients ($a$,$b$,$c$, and $d$) is determined. This is necessary in order to eliminate errors due to optical distortion. For the calculation of the transformation $Z_{v}^{n}$ the line pattern is projected onto a sprayed planar mirror that stands perpendicular to the $z$-axis. At several known $z$-positions within the entire measurement volume a camera image is acquired. Figure~\ref{fig6} (a) illustrates this procedure.
Since the values for $z$, $n$, $u$ and $v$ are known or can be determined, an overdetermined linear system of equations for each transformation $Z_{u}^{n}$  can be set up and solved by means of least squares. Finally, Equation~\ref{e_calib_z} can be used to calculate the depth values for arbitrary lines within the measurement volume. 

\subsection{$xy$-Calibration} 
\label{sec_calibration_lateral}

The lateral calibration allows the calculation of the lateral coordinates of the acquired 3D points. In the proposed approach the coefficients of the following two polynomial transformations are determined:
\begin{equation} 
	x = X(u,v,z) = a_{0} + \cdots + a_{17}u^3 + a_{18}v^3 + a_{19}z^3,
	\label{polynom_x}
\end{equation}
\begin{equation} 
	y = Y(u,v,z) = b_{0} + \cdots + b_{17}u^3 + b_{18}v^3 + b_{19}z^3,
	\label{polynom_y}
\end{equation}
where ($x$, $y$) is the desired lateral position in the 3D space and ($u$, $v$) the subpixel-precise position of the line in the camera image.
For this purpose, camera images of a planar surface with bright dots are acquired at the same $z$-positions which were used for the $z$-calibration. Figure~\ref{fig6} (b) demonstrates this approach. Since the positions of the dots are known in 3D space and can be localized in the camera images with high precision, one can again set up an over-determined linear system of equations to calculate the desired coefficients by means of least squares.

\subsection{Inverse Lateral Calibration} 
\label{sec_calibration_inverse}

Additionally, the color camera for the acquisition of texture information has to be calibrated. The procedure is basically the same as the described $xy$-calibration, only this time the inverse transformations have to be determined:

\begin{equation} 
	u = U(x,y,z) = a_{0} + \cdots + a_{17}x^3 + a_{18}y^3 + a_{19}z^3,
	\label{polynom_x_inv}
\end{equation}
\begin{equation} 
	v = V(x,y,z) = b_{0} + \cdots + b_{17}x^3 + b_{18}y^3 + b_{19}z^3,
	\label{polynom_y_inv}
\end{equation}
where ($x$, $y$, $z$) is the position of a point in 3D space and ($u$, $v$) the desired projection of the 3D point onto the pixel plane of the color camera. More details about the texture acquisition are given in Section~\ref{sec_texture}.

\subsection{Line Indexing}
\label{sec_indexing}

During a measurement process the calibration data is used to calculate a 3D view from each camera image. It is important that every single part of a line has to be labeled with the right line index $n$ to yield correct 3D data. The more lines are projected, the more difficult this will become (see Figure~\ref{fig7}).


\begin{Figure}
 \centering
 \includegraphics[width=\figurewidthsingle]{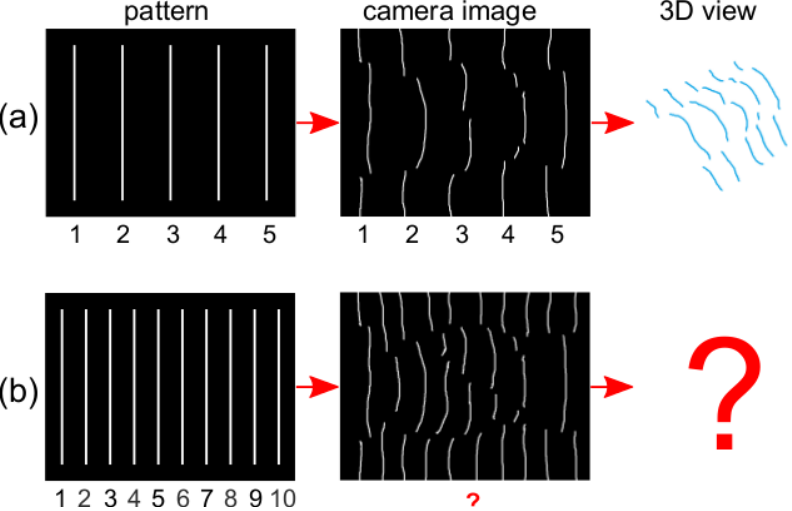}
 \captionof{figure}{Indexing: (a) Every line or part of a line has to be labeled with the right number to generate correct 3D data. (b) This becomes more difficult, the more lines are projected}
 \label{fig7} 
\end{Figure}

As we have already emphasized, no spatial or temporal encoding methods are used, since we do not want to decrease the qualitiy of the acquired 3D points by wasting bandwidth. Instead the basic idea for the indexing process is as follows: The sensor setup and the number of lines are chosen in a way that for objects inside the measurement volume every line is observed in a unique area on the camera image. That way, finding the right index for a line gets trivial. This works only as long as the line is observed within the measurement volume, otherwise lines can leave their unique area. Figure~\ref{fig8} illustrates this indexing approach.


\begin{Figure}
 \centering
 \includegraphics[width=\figurewidthsingle]{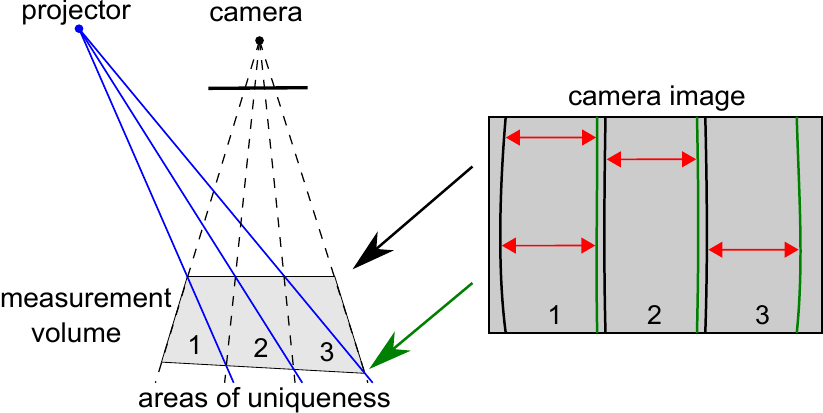}
 \captionof{figure}{Inside the measurement volume every line is observed in a unique area on the camera image}
 \label{fig8} 
\end{Figure}

Since it is hardly possible to guarantee that the object under test stays inside the measurement volume -- especially during a hand-guided measurement -- a further method is performed to detect and discard outliers. 
Due to the restricted depth of focus of the illumination and observation, the width of an observed line signal depends on its $z$-position. Specifically, outside the arranged measurement volume, the lines become considerably wider. A robust estimation for the line width can be easily obtained using the Gaussian interpolation described in Section~\ref{sec_localization}. 
Unfortunately, the broadening effect is happening continuously which makes it hard to identify outliers by their absolute line width.
Thus, a trick is applied. In the $z$-calibration not only transformations for the depth values but also transformations for the width of the lines are stored. During a measurement, the observed width of a line is normalized using the corresponding calibrated value. If a line is labeled with a wrong index, a false normalization factor is used and the normalized width differs more distinctly from the width of a correctly indexed line. 
As depicted in Figure~\ref{fig9}, outliers become even more distinguishable, if the illumination and observation are focused in front of the measurement volume. 
This way, a discrete step in the normalized width appears when a line is observed outside the measurement volume. Of course such an asymmetric focusing reduces the signal-to-noise ratio and leads to a slightly higher measurement uncertainty at the deeper part of the measurement volume.


\begin{Figure}
 \centering
 \includegraphics[width=\figurewidthsingle]{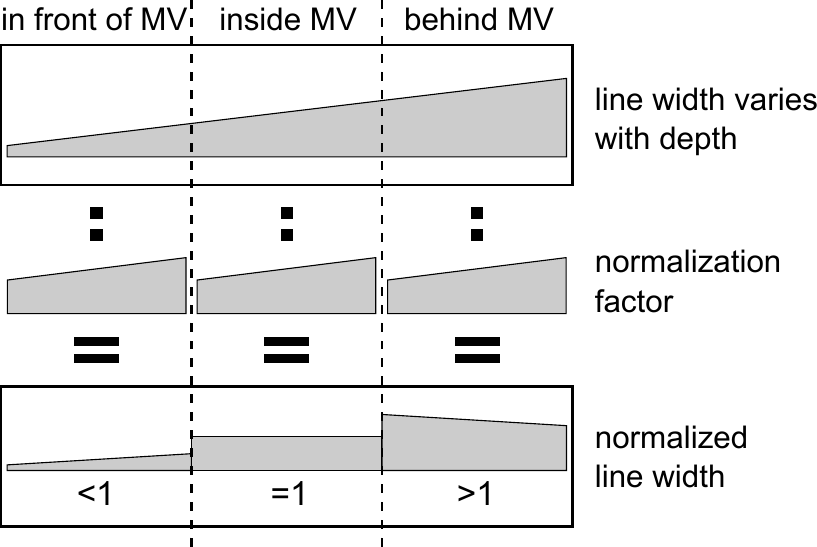}
 \captionof{figure}{The width of an observed line in a camera image depends on the depth of the measured object part. The normalized width remains constant as long as the line is observed inside the measurement volume (MV). If a line is observed behind or in front of the MV, its normalized width becomes significantly larger or smaller}
 \label{fig9} 
\end{Figure}

\section{Registration and Visualization}
\label{sec_registration}

This section describes the algorithms for a robust real-time registration and visualization of the acquired views. Moreover, an iterative global optimization for improving the registration result off-line is presented. The key ideas and concepts are summarized and indicated in a symbolic representation. Due to the limited space, details are avoided.

\subsection{Problem Assignment}
\label{sec_assignment}

\begin{figure*}
	\centering
	\includegraphics[width=\figurewidthdouble]{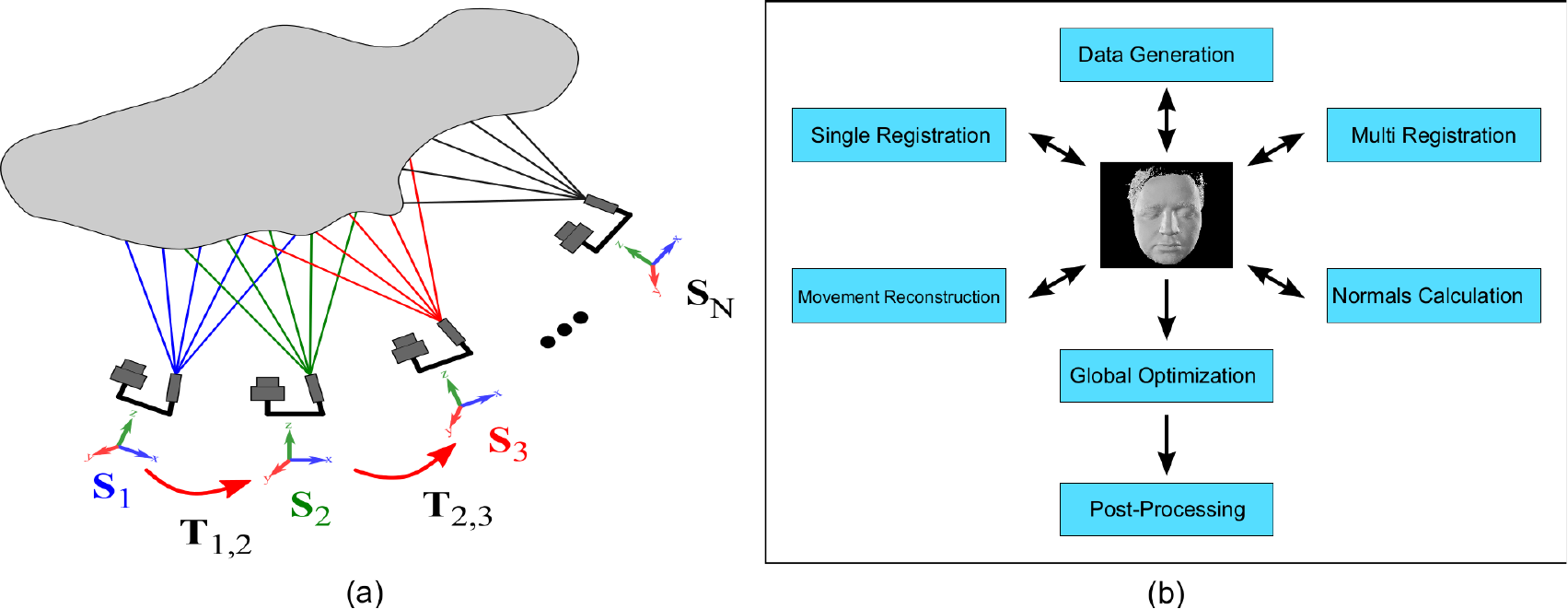}
	\caption{(a) The goal of the registration is to find the absolute sensor position and orientation $\mathbf{S}_i$ for every acquired view $i$. Therefore, each view is aligned to the previous one(s). (b) For an efficient real-time performance on a multi-core system the algorithms are separated in different modules which can run in parallel}
\label{fig10}      
\end{figure*}

First of all some general information and definitions are given for a better understanding of the concepts and algorithms. The calibration of the sensor establishes the so-called \textit{sensor coordinate system}, in which the acquired views are described. During the measurement procedure the sensor and the connected sensor coordinate system are moved. This means the coordinate systems of different views do not match. The registration determines for each view $i$ the transformation $\mathbf{S}_i$ that maps the particular view to the so-called \textit{world coordinate system}. The world coordinate system corresponds to the sensor coordinate system of the first acquired view of the specific measurement. In this case the transformation $\mathbf{S}_1$ of the first view is simply the identity matrix. Figure~\ref{fig10} (a) illustrates the registration task. For real-time application each view $i$ is registered to its predecessor $i-1$. This yields the transformation $\mathbf{T}_{i-1,i}$. The absolute transformation $\mathbf{S}_i$ is then given by:
\begin{equation} 
	\mathbf{S}_i = \mathbf{T}_{1,2} \cdot \mathbf{T}_{2,3} \cdot \mathbf{T}_{3,4} \cdot \ldots \cdot \mathbf{T}_{i-1,i}.
	\label{transformation}
\end{equation}

After the measurement a global optimization is used to register the views without using such a preferred order.  

\subsection{Modular Architecture}
In order to enable real-time capability, modern multi-core systems are exploited by separating the algorithms and workload in different modules which run in parallel. Figure~\ref{fig10} (b) gives an overview of the underlying algorithmic architecture of the modules. The number of parallel working modules can be easily adapted to the available computing power.

\subsection{Alternating Line Patterns}
\label{sec_alternating}

\begin{figure*}
	\centering
	\includegraphics[width=\figurewidthdouble]{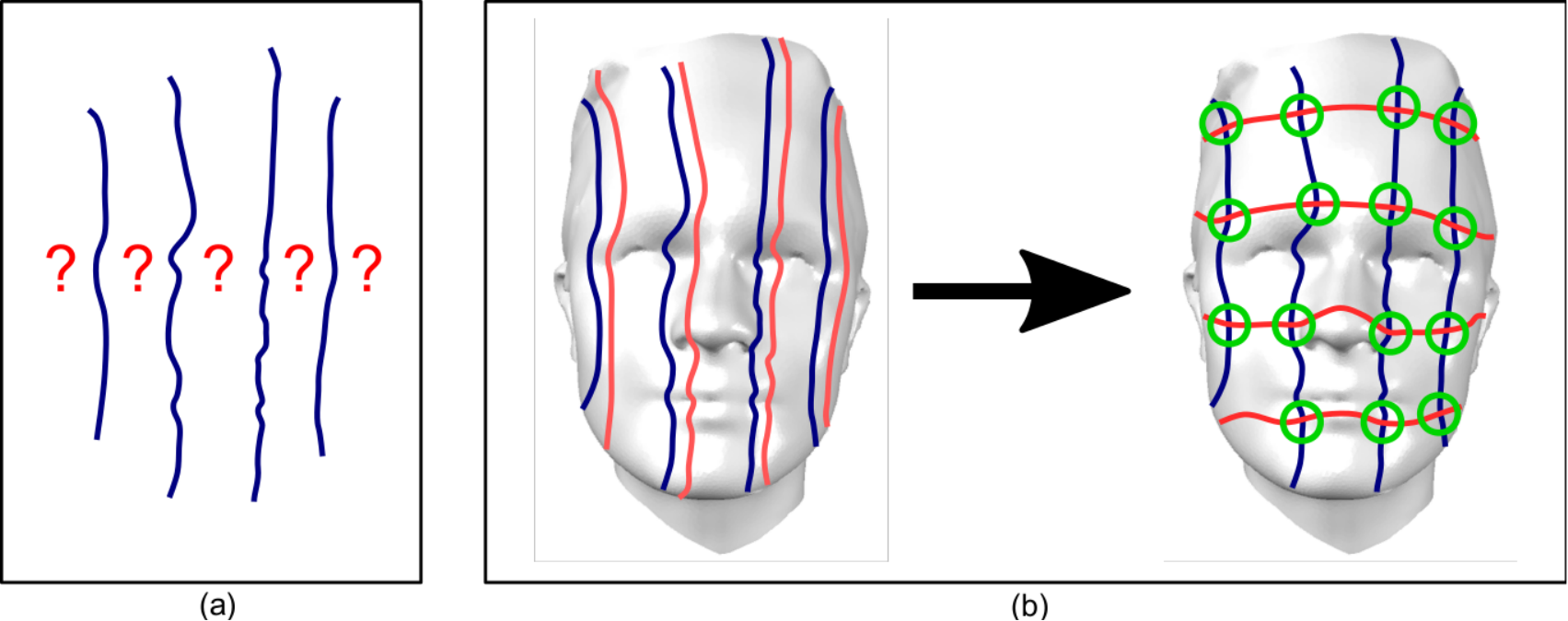}
	\caption{(a) 3D data is only acquired along the projected lines. Thus, no surface information between the lines is available for registration. (b) There are in general no or very few common points, if only views with vertical line profiles are acquired. The alternating acquisition of vertical and horizontal line profiles ensures the existence of several corresponding 3D points}
\label{fig11}      
\end{figure*}

As shown in Section~\ref{sec_sensor} the sensor acquires 3D points with high lateral and longitudinal resolution, but only along the projected multi-line pattern. Between the lines no data is obtained. Since, the sampling theorem is commonly not satisfied for a single view (because the lines are too far apart), there is no possibility to calculate reliable surface information in between. If the sensor is moved around the object and in each view 3D data is only acquired along the same line pattern, there is in general no or very few common information in two consecutive views (see Figure~\ref{fig11} a). For this reason we apply a trick: By alternate projection of vertical and horizontal lines and thereby an alternating acquisition of 3D views with vertical respectively horizontal line profiles the existence of corresponding points in consecutive views is guaranteed. Figure~\ref{fig11} (b) demonstrates this concept. The common 3D points can now be used to enable a robust and precise registration of the views.

\subsection{Corresponding Points}
\label{sec_corresponding}

%
%

In the proposed registration algorithms the closest points $(\mathbf{p}_{kl}, \mathbf{q}_{kl})$ between a vertical line profile $k$ and a horizontal line profile $l$ have to be found.
This can be done exactly by a brute-force search with a complexity of $N^2$, where $N$ is the number of points per line profile. Since the vertical and horizontal views are displaced only slightly to each other, the search space -- the number of considered points per view -- can be restricted significantly. Additionally, an hierarchical approach can be used to further accelerate the search.

\subsection{Single Registration}
\label{sec_fine}

The task of the single registration is to find the relative transformation between two 3D views. 
Under the eligible assumption that the sensor is moved only slightly between two consecutive views a fine registration can be applied directly in this case.
In the following a variant of an iterative closest point algorithm is described. It is specifically adapted to the sparse nature of the 3D views obtained by our proposed sensor system.

First, the acquired 3D data is smoothed along each line profile with a Gaussian filter in order to reduce the influence of noise on the registration process. Cubic splines are then used to interpolate 3D data between the discrete points along each line profile to allow a finer alignment of the views. 
Next, closest points are searched between the two views. Considering that view $V$ consists of vertical and view $H$ of horizontal line profiles, the existence of two corresponding points is assumed for every line profile pair $(k, l)$, where $k$ is the index of the vertical and $l$ of the horizontal line profile. For every such line profile pair the search described in Section~\ref{sec_corresponding} is used to determine the two corresponding points $(\mathbf{p}_{kl}, \mathbf{q}_{kl})$ and the distance $d_{kl}$ between them. Due to shadowing effects and the movement of the sensor not all line profile pairs $(k, l)$ intersect at corresponding points. Thus, point pairs $(\mathbf{p}_{kl}, \mathbf{q}_{kl})$ with a distance $d_{kl}$ greater than a dynamic threshold $d_{max}$ are discarded. 
For the remaining point pairs the mapping transformation by means of least squares is calculated. This process is now iterated until the changes are below a specific threshold or a maximum number of iterations is reached.

Now some further improvements are briefly described.
First, in the calculation of the mapping transformation the point pairs are weighted in order to increase the robustness of the registration process. This is done as follows: A scalar quantity that describes the variation of the $z$-values is calculated for every line profile. The higher the value the higher the variation. For each pair $(\mathbf{p}_{kl}, \mathbf{q}_{kl})$ the sum of the two involved values is used as a weighting factor.
By using only a random subset of line profile pairs respectively point pairs the iteration process can be accelerated considerably. Furthermore the search space for the point correspondences can be decreased step by step during the iterations.

\subsection{Multi-View Registration}
\label{sec_multi}

The goal of the multi-view module is to further refine the transformations obtained from the single registration described above. On a multi-core system it can be executed as a separate process to enable a more precise real-time alignment. The proposed method is based on a \textit{metaview} approach \citep{Chen92,Masuda96}. Each view is thereby not only registered to the previous view but to all previous views which behave together like one rigid view. The actual registration process is the same as for the fine registration but more line profile pairs respectively point pairs are used and the registration error is reduced.

Since the procedure is supposed to run in real time the search for valid line profile pairs -- line profiles that intersect each other -- has to be optimized. Thus, only previous views that overlap with the current view should be considered. For this purpose a three dimensional fixed-grid data structure that holds the positions (centers of mass) of all already registered views is constructed and updated in real time. Additionally, information about the basic orientation of the line profiles in the 3D space is stored for each view. Using this structure it is easy and fast to find views that are likely to overlap with the current view (see Figure~\ref{fig14}). 
 

\begin{Figure}
 \centering
 \includegraphics[width=\figurewidthsingle]{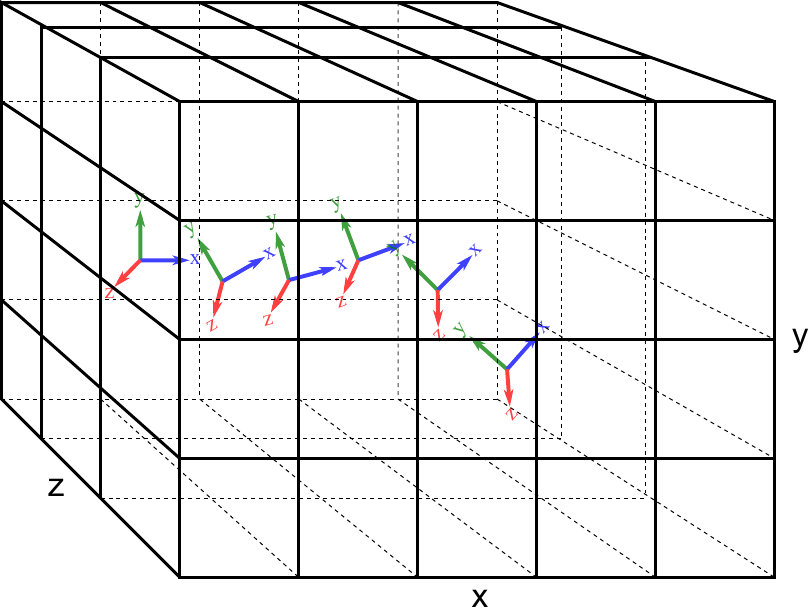}
 \captionof{figure}{A fixed-grid structure is filled with the absolute positions (center of mass) and orientations of the registered views. This allows a fast search for potential overlaps between views}
 \label{fig14} 
\end{Figure}

\subsection{Sensor Movement Reconstruction} 
\label{sec_reconstruction}

\begin{figure*}[!t]
	\centering
	\includegraphics[width=\figurewidthdouble]{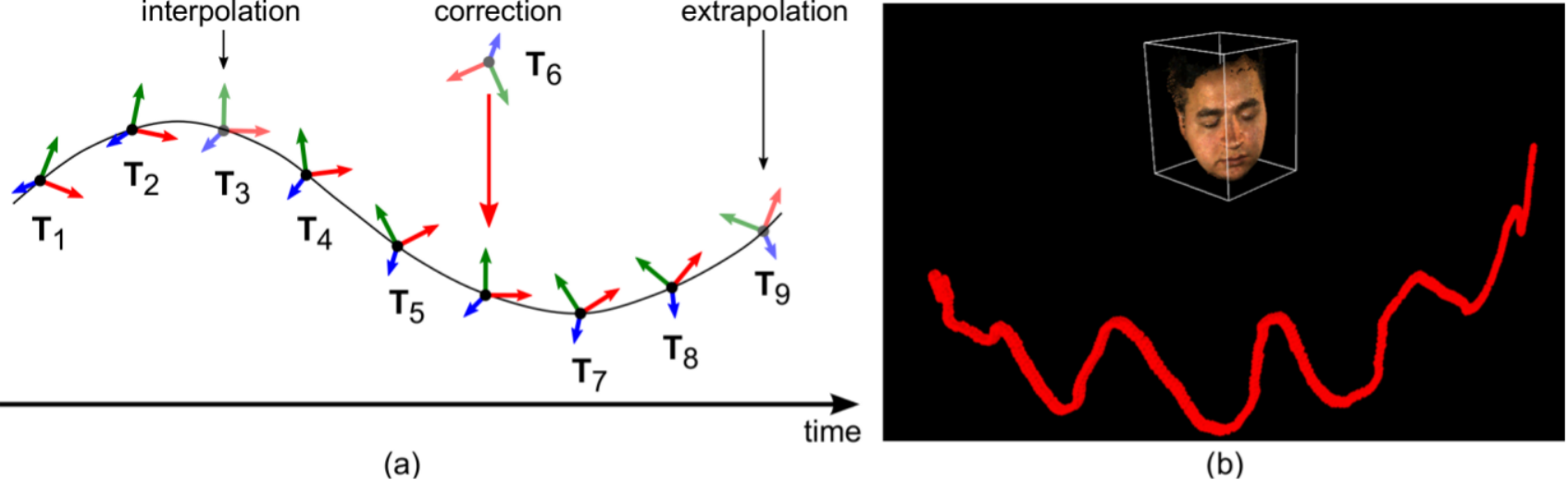}
	\caption{(a) Reconstruction of sensor movement: Outliers of the registration process are detected and eliminated. Transformations of skipped views can be obtained by interpolation. The next transformation can be estimated by extrapolation. (b) Reconstructed sensor path during a face measurement}
\label{fig15}  
\end{figure*}

Due to the sparse nature of the acquired views and little common information between them there is still a chance that the described registration algorithms get stuck in a local minimum or even fail. A method is implemented to further increase the robustness of the sensor by detecting and correcting outliers of the registration process. Under the assumption that the sensor is moved rather smoothly around the object the trajectory of the sensor in the six-dimensional transformation space is reconstructed. 
The relative transformations are represented as quaternions \citep{Horn87} and a fitting algorithm based on Random Sample Consensus \citep{Fischler87} is applied to find a best fitting curve under the presence of potential outliers. After the movement is reconstructed, the detected outliers are back projected onto the fitted curve and re-registered.
Besides the correction of outliers of the registration procedure, the reconstructed space curve can be used to obtain a more reliable start transformation 
for the above described registration method.
Furthermore, the efficiency of the registration -- especially for real-time application -- can be increased significantly by skipping the registration process for some views and ``guessing" their transformation by interpolation instead. Figure~\ref{fig15} summarizes these ideas.

\subsection{Calculation of Normals}
\label{sec_normals}

Now a method is described that allows a fast and accurate calculation of the normals of the measured points. It exploits the information obtained by the registration and can be executed on-line to enable a shaded visualization of the acquired point cloud in real time. Furthermore, the calculated normals allow an application of established techniques for point cloud triangulation like the \textit{Ball Pivoting Algorithm}~\citep{Bernardini99}.

\begin{figure*}
	\centering
	\includegraphics[width=\figurewidthdouble]{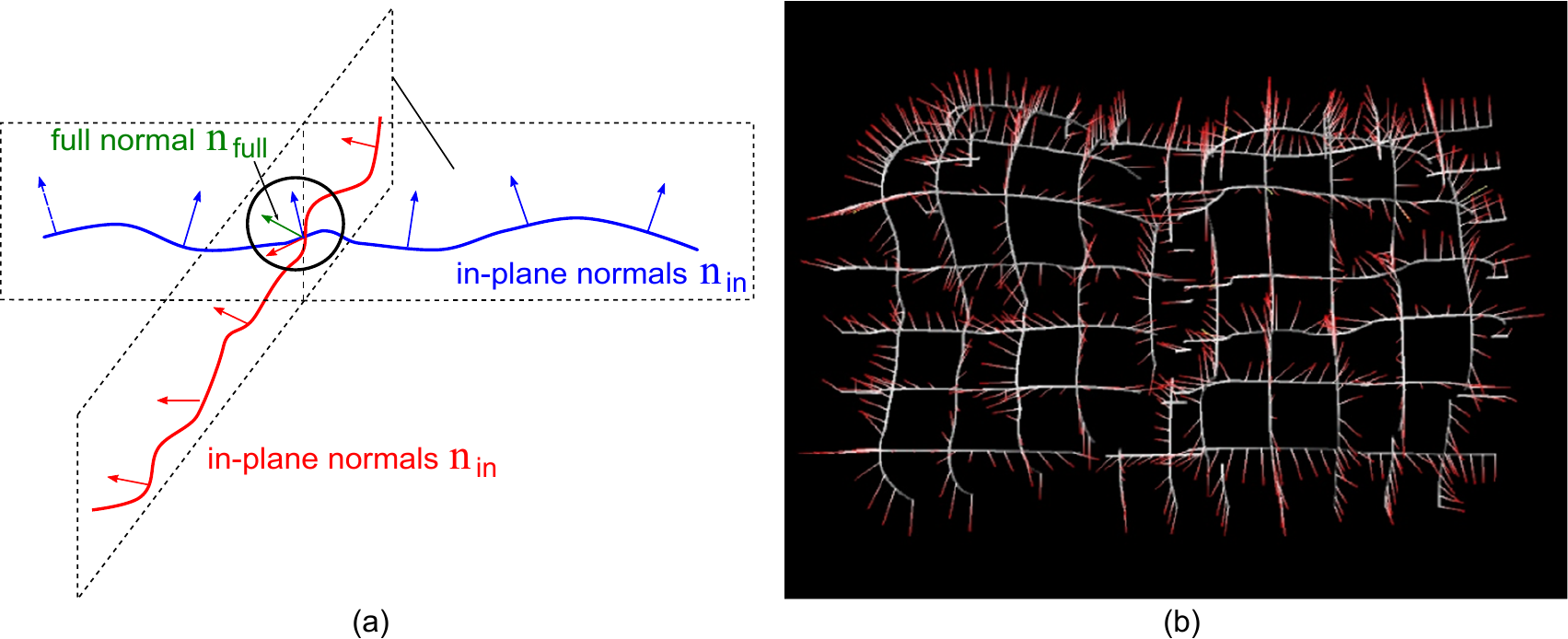}
	\caption{(a) Calculation the normal in the crossing point between two line profiles. (b) Two registered views with calculated and interpolated normals}
\label{fig16}      
\end{figure*}

Figure~\ref{fig16} illustrates the proposed procedure. First, for each point $\textbf{p}$ along each line profile a so-called \textit{in-plane-normal} $\mathbf{n}_{\mbox{\small in}}$, which lies inside the projected laser sheet, can be estimated using neighbouring points. A Gaussian filter can be applied to the 3D points beforehand to reduce the influence of noise. Furthermore, let $\mathbf{s}$ be the unit vector perpendicular to the plane in which the line profile lies. The actual normal $\mathbf{n}_{\mbox{\small full}}$ must lie in the plane perpendicular to the (unit) vector $\mathbf{o} = \mathbf{n}_{\mbox{\small in}} \times \mathbf{s}$.

If a line profile pair $(k,l)$ was used in the registration process, then there is also a corresponding point pair $(\textbf{p}^{k}, \mathbf{p}^{l})$ which marks the intersection between the two line profiles (see Section~\ref{sec_fine}). In case of perfect data and registration, $\textbf{p}^{k}$ and $\mathbf{p}^{l}$ would of course be equal and describe the crossing position between the two line profiles. 
Thus, the actual normal at the crossing position can be easily calculated by the intersection of the planes described by the vectors $\mathbf{o}^{k}$ and $\mathbf{o}^{l}$:

\begin{equation}
\label{full_normal}
	\textbf{n} 	= \mathbf{o}^{k} \times \mathbf{o}^{l} 
							= (\mathbf{n}^{k}_{\mbox{\small in}} \times \mathbf{s}^{k}) \times (\mathbf{n}^{l}_{\mbox{\small in}} \times \mathbf{s}^{l}).
\end{equation} 

This calculation can only be performed for points that are matched during the registration process. A trivial approach to determine the normals for the other points is to simply interpolate respectively extrapolate the normals along each line profile. A more sophisticated and precise method is the following: Instead of a direct interpolation between the calculated normals the vectors $\mathbf{o}$ of the particular corresponding points are interpolated along the line profiles and then again equation~\ref{full_normal} is applied to calculate the normals.

\subsection{Preview and Continuation Mode}
\label{sec_reentry}

For a user-friendly positioning of the sensor to the object under test (or  vice versa), the so-called \textit{preview mode} is started. In this mode the projection of the line and the generation of 3D views is executed as during a measurement, but the registration modules are still turned off. The last two views (one horizontal and one vertical) are displayed on the screen (see Figure~\ref{fig17} a). This allows the operator to ensure that the object is inside the measurement volume. As soon as this is the case, the operator can start the actual measurement by turning on the registration.


\begin{Figure}
 \centering
 \includegraphics[width=\figurewidthsingle]{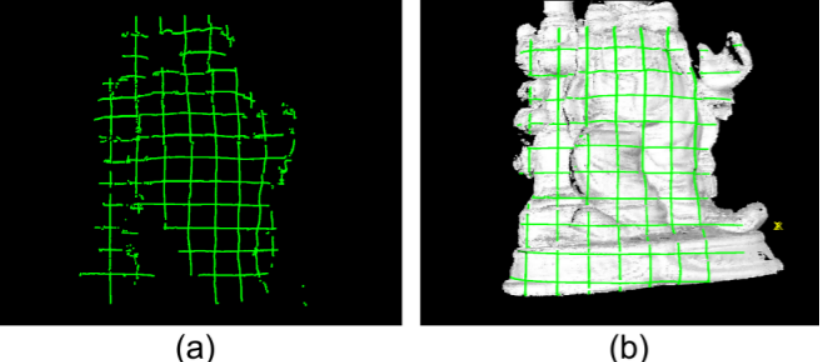}
 \captionof{figure}{(a) The \textit{preview mode} allows the operator to position the sensor or object. (b) If the measurement process was interrupted, the \textit{continuation mode} is started.  Two appropriate views for a re-entry are determined and highlighted on the screen}
 \label{fig17} 
\end{Figure}

If -- during a measurement -- the object under test moves out of the measurement volume or if too few lines are observed, the consecutive registration may fail. Hence, the following approach is applied to increase the robustness of the measurement procedure:
Let $N-1$ be the number of the last successfully registered view. If the registration of view number $N$ to view number $N-1$ fails, the measurement is not stopped immediately. Instead the registration of the next $k$ views $N+1, N+2,..., N+k$ to view number $N-1$ is attempted as well, where $k$ is typically a number between 15 and 30.  If this also fails, the measurement is interrupted and the so-called \textit{continuation mode} is started to allow a re-entry:

First, for each of the last $l$ successfully registered view pairs $(N-1,N-2), (N-2,N-3),...,(N-l,N-l-1) $ a rating is calculated that describes the applicability for the registration process of the views. Tests have shown that the sum of line profiles, or the sum of points are appropriate ratings. Now, the view pair with the highest rating is chosen and the particular views are highlighted in the real-time feedback. The operator can now move the sensor (or the object) to the desired position and the registration continues. Figure~\ref{fig17} (b) shows an example.

\subsection{Global Optimization}
\label{sec_global}

After the measurement a global iterative optimization process is applied to refine the registration result without real-time demand. The method is an adaptation of the multi-view registration described in Section~\ref{sec_multi}, but now the order in which the views were acquired is no longer of relevance.
Instead one view $V$ is picked randomly and all overlapping views are determined. These overlapping views are now treated as one rigid metaview $U$ and the view $V$ is then registered to this metaview $U$ using the fine registration of Section~\ref{sec_fine}.
Next, the procedure is performed with another randomly picked view and so on. 
If all views were re-registered once the whole process is repeated until the changes of the transformations are below a threshold.

\section{Post-Processing}
\label{sec_post}

This section describes two procedures that are applied after the measurement and registration. One for the reduction of the acquired point cloud and one for the mapping of texture images onto the data.

\subsection{Data Reduction}
\label{sec_reduction}

As already described, 3D points are acquired by scanning the object with a multi-line sensor at 30 frames per second.
This process leads to very dense or even redundant data. Different lines pass the same part of the object. The operator may even have re-scanned an already measured part by moving the sensor back and forth. Figure~\ref{fig18} shows an example. Due to the unavoidable, statistical measurement uncertainty of the acquired 3D points the resulting surface description displays a certain ``thickness". 


\begin{Figure}
 \centering
 \includegraphics[width=\figurewidthsingle]{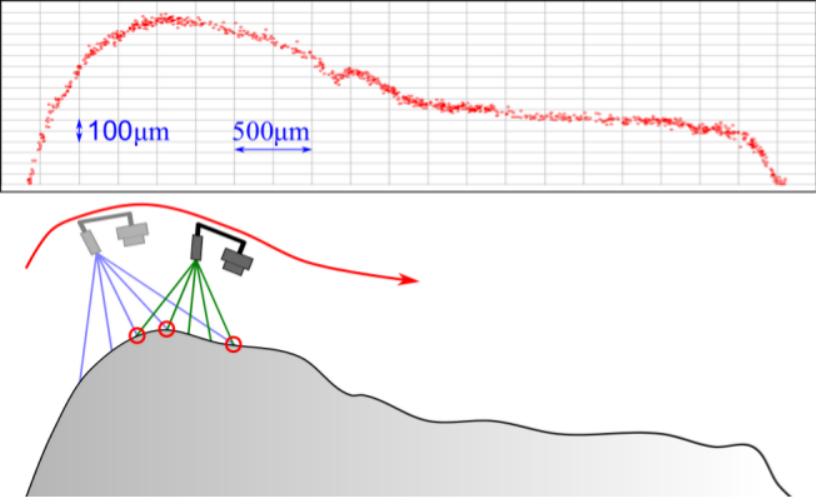}
 \captionof{figure}{While the sensor is moved along the object different lines scan over the same part of the object. This results in a ``thickness" of the acquired surface}
 \label{fig18} 
\end{Figure}

\begin{figure*}[!t]
	\centering
	\includegraphics[width=\figurewidthdouble]{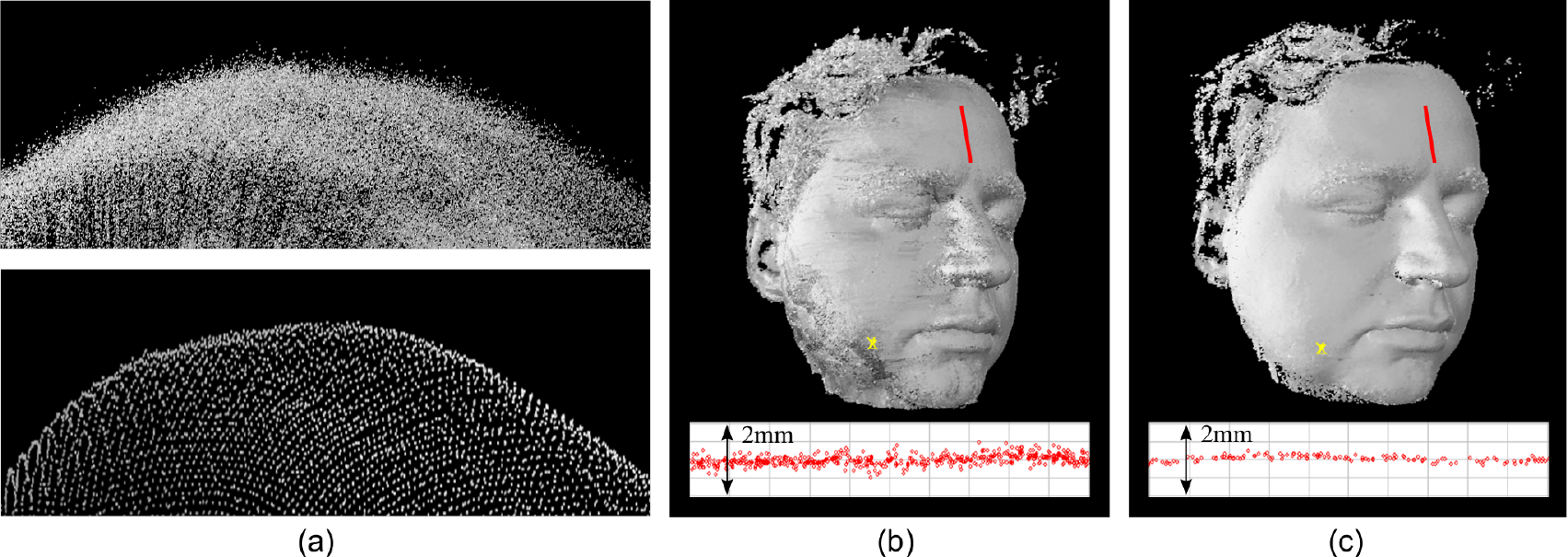}
	\caption{(a) Zoomed view of a point cloud before and after data reduction. Middle: Face measurement before data reduction. (b)  Face measurement after data reduction. Since each resolution cell is filtered individually, no lateral information is lost}
\label{fig19}      
\end{figure*}

By merging and averaging the data in the direction of the surface normals it is possible to thin out the point cloud and reduce the noise without losing lateral resolution. That means small details are not smoothed and are still visible in the final data set. This is accomplished by applying the following steps: An (arbitrary) point $\mathbf{p}$ is chosen and a small cell $C^{\mathbf{p}}$ around it elongated in the direction of the normal vector. All points that are enclosed by the cell $C^{\mathbf{p}}$ are merged to one single average point $\mathbf{p}_{\mbox{\small avg}}$. The corresponding normal vector $\mathbf{n}_{\mbox{\small avg}}$ is also obtained by averaging the normal vectors of the points enclosed by the cell $C^{\mathbf{p}}$. Now these steps are repeated until all points lie in separate cells. Figure~\ref{fig19} displays point clouds before and after the data reduction. 

\subsection{Texture Acquisition and Mapping}
\label{sec_texture}

\begin{figure*}
	\centering
	\includegraphics[width=\figurewidthdouble]{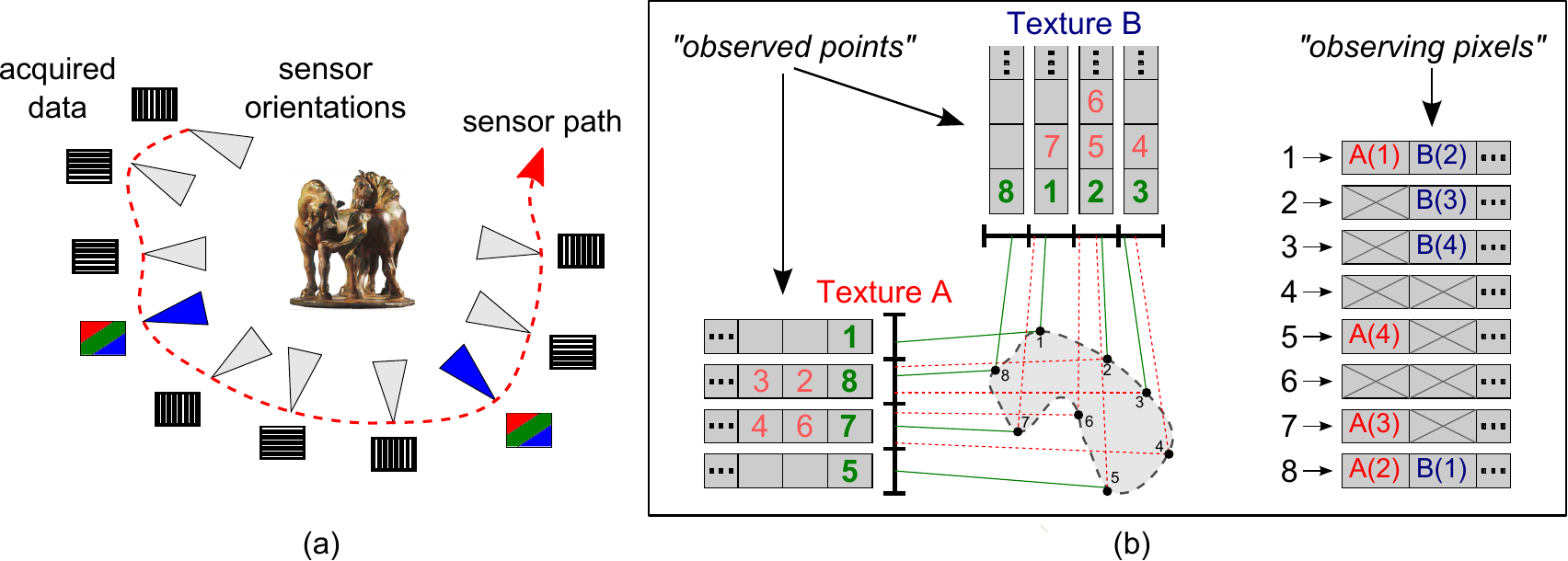}
	\caption{(a) The absolute positions of the sensor while the texture images were acquired are obtained by interpolation between the known positions. (b) Basic concept of the texture mapping. First, for each texture pixel a list of all potentially observed 3D points is filled (\textit{observed points}). Second, for each point a list of all texture pixels in which it is actual observed is determined under consideration of shadowing effects (\textit{observing pixels})}
\label{fig20}      
\end{figure*}

The task of the proposed approach is to ``colorize" the acquired point cloud. In principle it is possible to use a color camera for both 3D acquisition and texture acquisition, but a higher measurement precision is achieved with a black and white camera. For this reason an additional color camera and a flash light is integrated in the sensor system to obtain texture information during the measurement process. Periodically -- for example every second -- the 3D acquisition is briefly interrupted and a color image is captured. Of course the measurement procedure can still be performed as usually. The color camera has to be calibrated in order to allow a mapping from the sensor coordinate system to the pixel plane of the camera (see Section~\ref{sec_calibration_inverse}).  
Moreover, the position and orientation of the sensor in world coordinates (while a texture image was acquired) has to be determined. This is done by interpolation between the sensor transformations before and after the texture acquisition, as illustrated in Figure~\ref{fig20} (a). This allows to transform points given in world coordinates to the particular sensor coordinates system. 
The basic idea for texturing the acquired point cloud is illustrated in Figure~\ref{fig20} (b).

First, for each texture image $T$ every single 3D point $\mathbf{p}$ is transformed to the corresponding sensor coordinate system and then mapped onto the image plane of the texture. In this way, each 3D point yields a list called \textit{observing pixels} containing all pixels, by which the 3D point is potentially observed and each texture pixel yields a list called \textit{observed points} comprising all 3D points, which lie on its ray of sight. 
Shadowing effects are considered by applying a $z$-buffering technique: Only the points which are closest to the pixels are actually observed by the particular pixel. Both lists, \textit{observing pixels} and \textit{observed points}, are updated to satisfy this condition.

Now, for each 3D point a list with all valid textures respectively pixels is obtained. 
Due to noise, due to errors in the position interpolation, and mainly due to illumination variations under different angles, the observing pixels for one 3D point do not show the same color and intensity values. 
Measurement examples have shown that choosing the average color and highest occurring intensity value delivers the best visual result. Figure~\ref{fig21} depicts an example. A quantitative evaluation is pending.


\begin{Figure}
 \centering
 \includegraphics[width=\figurewidthsingle]{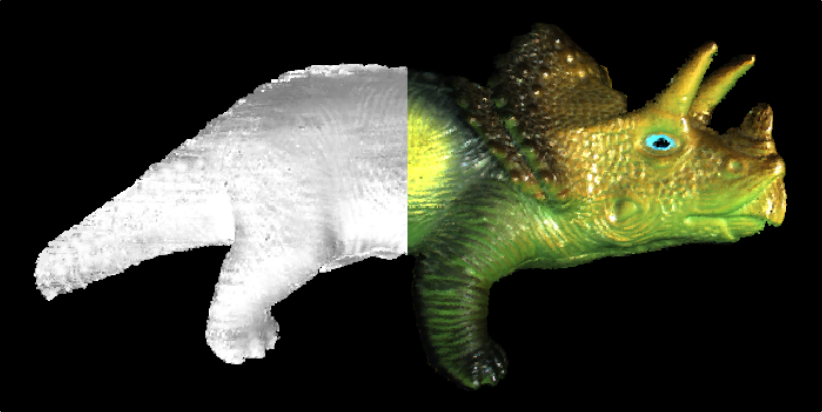}
 \captionof{figure}{Point cloud half textured, half non-textured}
 \label{fig21} 
\end{Figure}

\section{Results}
\label{sec_results}

In this section, first the precision of the registration is evaluated with the help of a simulation toolbox. Then, some measurement examples are presented to demonstrate the performance and applicability of the sensor principle.

\subsection{Simulation} 
\label{sec_simulation}

\begin{figure*}[!t]
	\centering
	\includegraphics[width=\figurewidthdouble]{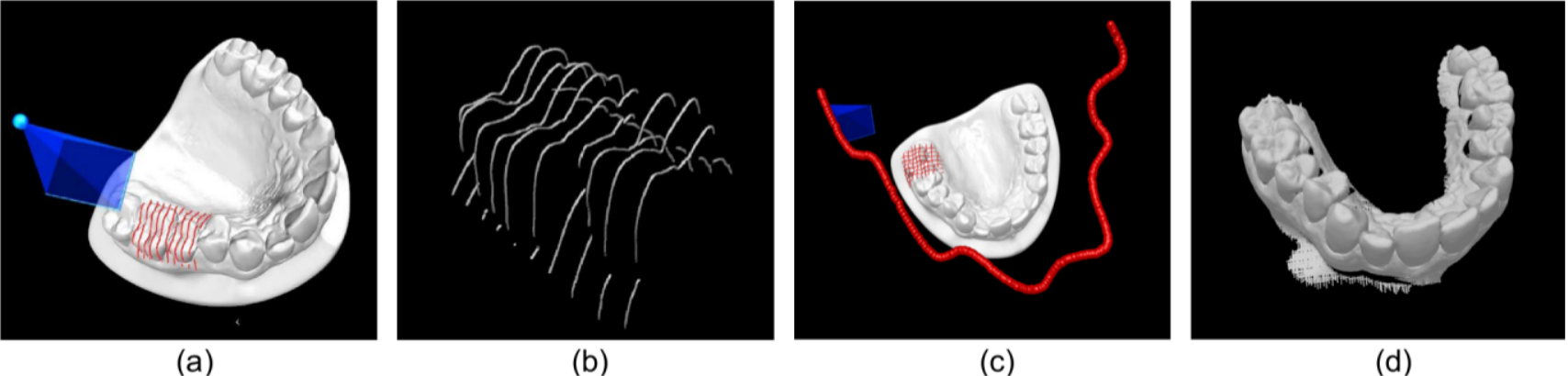}
	\caption{Simulation: A virtual sensor (a) acquires views (b) along a defined sensor path (c) and registers them to a dense point cloud (d)}
\label{fig22}      
\end{figure*}

For the evaluation of the registration error a simulation toolbox has been developed. Figure~\ref{fig22} illustrates the simulation process. A virtual sensor acquires 3D views from a 3D model. By defining an arbitrary sensor path around the object one can simulate a complete measurement. Furthermore, noise extracted from real measurements of a sprayed mirror can be added to the generated series of views before the registration algorithms are applied.

Since we always know the exact position of the virtual sensor, we can compare the ideal point cloud with the registration result point by point. Additionally, the added noise can be removed after the registration was applied, in order to calculate the pure registration error. 
If no noise is added to the data at all -- that means the registration is performed on ideal data -- the registration error only depends on the discrete nature of the acquired data. Since the sampling theorem along the line profiles is in general not perfectly satisfied, a marginal registration error remains. 

In the evaluated simulation process a full dental cast was virtually scanned by acquiring 1000 views. Table~\ref{tab_stats} displays the resulting statistical information. The variance of the added noise is approximately 30$\upmu$m. The total average point-to-point error directly after the real-time registration is about 300$\upmu$m. After the global optimization the error is reduced to 32$\upmu$m. If the noise is removed after the registration, the point-to-point error is approximately 24$\upmu$m, which is clearly below the measurement uncertainty within a single view. Finally, applying the registration to perfect data without noise leads to an error of less than 5$\upmu$m. \\
The results prove that the proposed sensor principle and algorithms allow a highly precise registration of the acquired views. 

\begin{Table}	    
	\centering
	\captionof{table}{Statistical results of the registration on data acquired by a simulated sensor (Intel Core 2 Quad CPU Q9550 2.83GHz)}
	\label{tab_stats}	

	\begin{tabular}{|p{5.0cm}|p{2.5cm}|} 
		\hline \noalign{\smallskip}
			
		Measurement volume 									& 20$\times$15$\times$15mm$^3$  \\	
		Vertical lines 											& 10 									\\	
		Horizontal lines										& 7 									\\	
		Model size (dental cast) 						& 80$\times$60$\times$30mm$^3$  \\	
		Acquired 3D views 									& 1000  							\\
		Acquired 3D points									& 4,500,000  					\\
		Added (real) noise									& 30$\upmu$m 						\\		
		Real-time error (noise included)		& 300$\upmu$m						\\
		Final error	(noise included) 				&	32$\upmu$m 					\\
		Final error (noise removed) 				& 24$\upmu$m   				\\
		Final error (ideal data) 						&  5$\upmu$m   				\\
			
		\noalign{\smallskip}\hline	
	\end{tabular}	
\end{Table}

\subsection{Measurement Examples} 
\label{sec_examples}

In the following, two measurement examples acquired by a sensor designed for mid-sized objects -- like small sculptures or body parts -- are presented. The measurement volume is about 200mm $\times$ 150mm $\times$ 100mm, the triangulation angle is $\Theta = 7^{\circ}$, and the physical measurement uncertainty is below 120$\upmu$m. See Section~\ref{sec_optimization} and \citep{Willomitzer10} for further information. 

\begin{figure*}[!t]
	\centering
	\includegraphics[width=\figurewidthdouble]{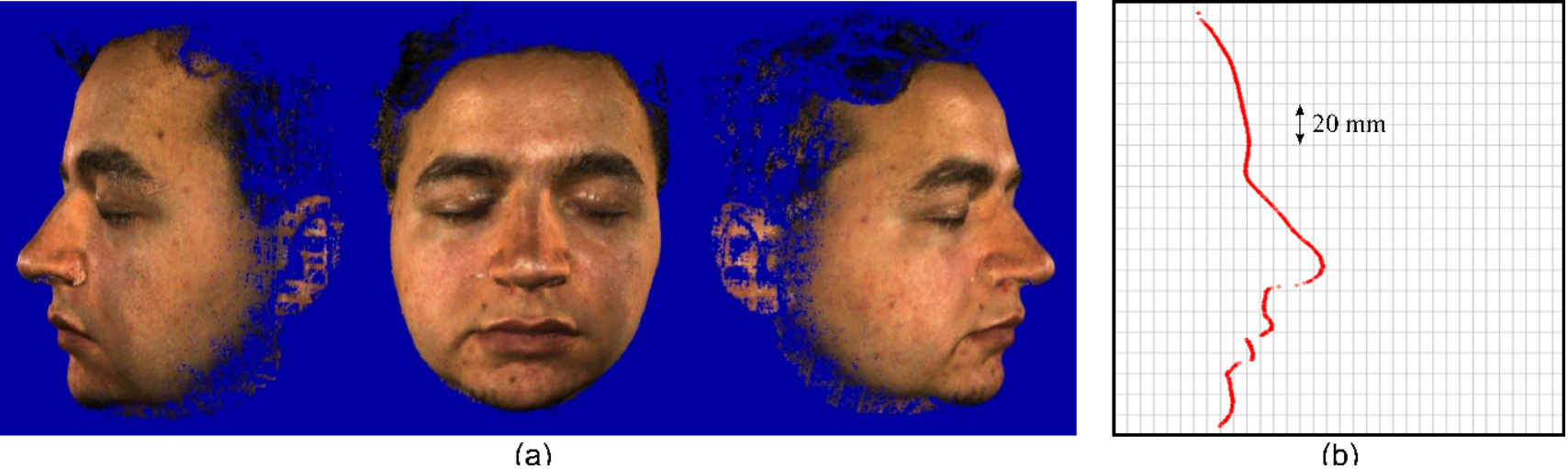}
	\caption{(a) $360^{\circ}$ measurement of small sculpture. (b) Resulting point cloud. (c) Cross section through point cloud}
\label{fig23}      
\end{figure*}

The first example demonstrates the flexibility and easy handling of the sensor. Figure~\ref{fig23} displays the result of a 360 degree measurement of a small sculpture. During the measurement the the sculpture was turned around by hand, while the sensor stood still. The cross section shows that after the full turn of the sculpture a closed point cloud is received.

\begin{figure*}
	\centering
	\includegraphics[width=\figurewidthdouble]{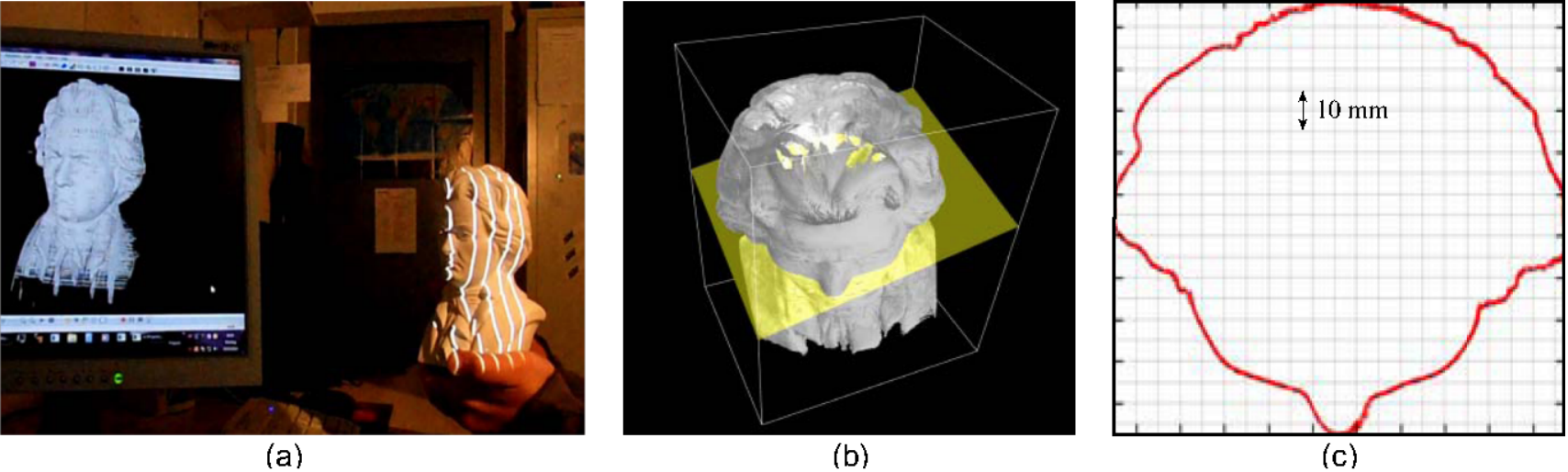}
	\caption{(a) Textured face measurement from different angles. (b) Cross section through point cloud}
\label{fig24}      
\end{figure*}

The second example displays a measurement of the face of the first author. This time the sensor was moved around the face and a 180 degree point cloud was acquired. During the measurement time of approximately 15 seconds the face was not allowed to perform any intense non-rigid movements in order to obtain consistent 3D data. 
Additionally to the pure 3D acquisition a color texture was captured and mapped onto the data as described in Section~\ref{sec_texture}. Figure~\ref{fig24} displays the resulting textured point cloud from different angles and a cross section through it. 

\section{Conclusion}
\label{sec_conclusion}

We introduced a sensor and an algorithmic pipeline which enables a new measurement principle: Flying Triangulation. A continuous stream of sparse 3D views is acquired by a hand-guided sensor based on light sectioning. The algorithms are able to merge the views to a dense point cloud and visualize the running result in real time as a useful feedback for the operator. No tracking system restricts the movement of the sensor. Furthermore, colored images are acquired and mapped onto the point cloud. 
The main characteristics of the presented sensor system are the following: Instead of full-field data, dense and precise data along separate lines is acquired. Since each acquired 3D point provides a high depth and high lateral resolution, small details can be resolved. By the alternate projection of vertical and horizontal line patterns, a robust alignment of the data is enabled. It is shown that this approach leads to a low error propagation along the series of 3D views. 
Both coarse and fine registration specifically suited for this kind of data are introduced and evaluated. Additionally, a reconstruction of the sensor movement allows the detection and elimination of outliers and thereby improves the robustness. Due to the scanning nature of the sensor, very dense or even redundant data is acquired during a measurement. By applying proper data reduction, noise is decreased without loss of lateral information. Simulations and real measurements demonstrate the flexibility and functionality of the presented sensor and algorithms.

In the following, the most important remaining challenges and problems of the proposed sensor system are explained.
The success of the registration depends on the information provided by the sparse 3D views. If parts of the object do not have sufficient structure or too few lines deliver 3D data, the registration is prone to fail. 
There are in principle three ways to overcome this difficulty: 
First, the registration has to be adapted to objects which only have little structure and contain many flat parts. 
Second, the number of lines per view should be increased significantly to guarantee the existence of sufficient 3D information per view even if the object is relatively small. 
Third, the size of the measurement volume has to be increased in order to ensure that the object under test stays inside the volume while measuring.
For both the second and third aspect it is necessary to develop a more sophisticated indexing without being restricted to the areas of uniqueness that are described in Section~\ref{sec_indexing}.
Besides the improvement of the robustness of the sensor, there are several additional challenges: The data reduction and the texture mapping should be integrated in the real-time architecture. Furthermore, a proper triangulation of the resulting point cloud is desired.



\end{multicols}

\end{document}